\documentclass[preprint,12pt,authoryear]{elsarticle}

\usepackage{amssymb}

\journal{Advances in Space Research (ASR)}

\newcommand{\aj}{AJ}
\newcommand{\apj}{ApJ}
 
\newcommand{\apjs}{ApJS} 
\newcommand{\aap}{A\&A}
\newcommand{\mnras}{MNRAS}
\newcommand{\pasp}{PASP}
\newcommand{\apss}{Ap\&SS}

\newcommand{\Msun}{M_{\odot}}

\newcommand{\Msec}{M_{\mathrm{sec}}}
\newcommand{\Rsec}{R_{\mathrm{sec}}}

\newcommand{\Porb}{P_{\mathrm{orb}}}
\newcommand{\Mwd}{M_{\mathrm{WD}}}

\newcommand{\Msecdot}{\dot{M}{_\mathrm{sec}}}

\newcommand{\Jcamldot}{\dot{J}_{\mathrm{CAML}}}
\newcommand{\Jsysdot}{\dot{J}_{\mathrm{sys}}}
\newcommand{\Jdot}{\dot{J}}

\newcommand{\lppr}{\stackrel{<}{\scriptstyle \sim}}
\newcommand{\lappr}{\raisebox{-0.4ex}{$\lppr$}}

\begin{document}

\begin{frontmatter}



\title{Cataclysmic variable evolution and the white dwarf mass problem: A Review}

\author[label1]{Monica Zorotovic}
\author[label1,label2]{Matthias R. Schreiber}
\address[label1]{Instituto de F{\'i}sica y Astronom{\'i}a, Universidad de Valpara\'iso, Av. Gran Breta\~na 1111, Valpara\'iso, Chile}
\address[label2]{Millenium Nucleus for Planet Formation, Universidad de Valpara{\'i}so, Valpara{\'i}so 2360102, Chile}

\begin{abstract}
Although the theory of cataclysmic variable (CV) evolution is able to explain several observational aspects, strong discrepancies have existed for decades 
between observations and theoretical predictions of the orbital period distribution, the location of the minimum period, and the space density of CVs. 
Moreover, it has been shown in the last decade that the average white dwarf (WD) mass observed in CVs is significantly higher than the average mass 
in single WDs or in detached progenitors of CVs, and that there is an absence of helium-core WDs in CVs 
which is not observed in their immediate detached progenitors. 
This highly motivated us to revise the theory of CV formation and evolution. 
A new empirical model for angular momentum loss in CVs was developed in order to explain the high average WD mass observed and the absence of 
systems with helium-core WDs. This model seems to help, at the same time, with all of the above mentioned disagreements between theory 
and observations. Moreover, it also provides us with a very likely explanation for the existence of low-mass WDs without a companion.
Here we will review the standard model for CV evolution and the disagreements that have existed for decades between simulations and observations with their possible 
solutions and/or improvements. We will also summarize the recently confirmed disagreement related to the average WD mass and the fraction of helium-core WDs among CVs,
as well as the development of an empirical model that allows us to solve all the disagreements, discussing the physics that could be involved. 

\end{abstract}


\end{frontmatter}


\section{Introduction}
\label{intro}

Cataclysmic variables (CVs) are semi-detached short period binaries in which a white dwarf (WD) primary
stably accretes mass from a low-mass companion, hereafter the donor or secondary star \citep[see][for a comprehensive review]{warner95-1}. 
The donor is typically an unevolved low-mass main-sequence (MS) star, although systems at orbital periods below $\sim 1$\,h or 
above $\sim 5-6$\,h can have a degenerate companion or a slightly evolved donor, respectively
\citep[e.g.,][]{goliasch+nelson15-1,kalomenietal16-1}.

One of the most striking observational properties of the population of CVs is the statistically significant deficit of 
systems in the orbital period range of $2-3$\,h, known as the ``orbital period gap''. In order to explain this observational feature, 
the ``disrupted magnetic braking'' model for angular momentum loss in CVs was proposed by \citet{rappaportetal83-1} and \citet{spruit+ritter83-1}. 
This model assumes that CVs lose angular momentum as a result of two mechanisms: gravitational radiation \citep{paczynski+sienkiewicz81-1}
and wind magnetic braking \citep{verbunt+zwaan81-1}. The latter is much stronger than gravitational radiation when the donor still has a 
radiative core, but it would become inefficient when the donor becomes fully convective, due to a sharp decline in the magnetic dynamo.
This is expected to occur in MS stars with $\sim 0.3\Msun$, which corresponds to an orbital period of $\Porb \sim 3$ h 
for a Roche Lobe filling MS star with a WD primary. 
An alternative explanation for the strong decline in the efficiency of magnetic braking close to the upper edge of the period gap
was recently explored by \citet{garraffoetal18-1}, motivated by X-ray observational evidence that suggests that there is no change in the surface 
magnetic activity near the fully convective boundary for MS stars, which is needed to cause a change in the dynamo generation rate of magnetic flux.
The authors were able to explain the gap based only on the complexity of the surface magnetic field of the donor star. 
However, although the origin of the reduction in the efficiency of magnetic braking suggest by \citet{garraffoetal18-1} is different 
than the dynamo explanation given by \citet{rappaportetal83-1} and \citet{spruit+ritter83-1}, the effects in the rate of angular momentum 
loss is similar, and the disrupted magnetic braking model is still valid.
The strong reduction in the efficiency of magnetic braking would cause a sharp reduction in the mass transfer rate allowing the donor, 
which was previously slightly extended as a result of mass transfer, to relax towards its equilibrium radius, 
temporarily ceasing mass transfer until it fills its Roche lobe again at $\Porb \sim 2$\,h. 
The system will continue to evolve towards shorter orbital periods through the loss of angular momentum by gravitational radiation only,
until the donor becomes degenerate at $\Porb \sim 60-80$\,min. The response of the degenerated donor to the loss of mass, i.e. expansion,
will cause CVs to evolve back towards larger orbital periods, which predicts the existence of a minimum orbital period
\citep{rappaportetal82-1,paczynski+sienkiewicz83-1}. CVs at shorter periods (e.g., AM CVns) are 
thought to contain helium-rich degenerate donors instead of MS donor stars, whose mass-radius relation allows them 
to reach periods as short as $\sim 5-6$\,min \citep[see e.g.,][and references therein]{kalomenietal16-1}.

A still open question related to the evolution of CVs is whether the mass of the WD can grow as a consequence of mass transfer.
According to standard theories for nova outbursts, the mass that is ejected during an eruption is similar to the mass accreted 
between them, implying that the mass of the WD remains nearly constant during the CV phase \citep{prialnik86-1,prialnik+kovetz95-1,yaronetal05-1}.
Only the more massive WDs accreting at very high rates are expected to grow in mass \citep[e.g.][]{hillmanetal15-1}.
Observationally, \citet{mcallisteretal19-1} found no evidence of mass growth in CV WDs, by comparing the average WD mass in CVs above and below the
period gap, but they also mentioned that more systems with precise WD masses above the gap are needed to derive a definitive answer. 

The disrupted magnetic braking model became the standard model of CV evolution for almost three decades and is supported by 
observational evidence coming from CVs \citep{townsley+bildsten03-1,knigge06-1,townsley+gaensicke09-1}, as well as from their 
detached progenitors \citep{schreiberetal10-1}, and even from single stars \citep{bouvier07-1,reiners+basri08-1}. However, there are 
also several disagreements between the observations and the theoretical predictions perceived since the first binary 
population models of CVs \citep[e.g.,][]{dekool92-1,kolb93-1,politano96-1}, 
which have persisted in binary population studies conducted in the subsequent decades 
\citep[e.g.,][]{howelletal01-1,goliasch+nelson15-1}.

Here we will review the main disagreements between the theoretical predictions (based on this model) and the observations of CVs, 
as well as the latest advances in the theory of CV evolution developed in recent years.
Section\,\ref{sec2} summarizes the main disagreements that have existed for decades, and their possible solutions and/or improvements.
In Section\,\ref{sec3}, we present a new disagreement evidenced in the last decade, which probably became the most important 
problem in the theory of CV evolution, known as the ``WD mass problem''. Section\,\ref{sec4} explains a new evolutionary model for CVs, 
which was empirically derived in order to solve the WD mass problem. Independent tests for this model are presented in Section\,\ref{sec5}
while Section\,\ref{sec6} discusses a possible physical explanation behind this new approach. 
A summary is presented in Section\,\ref{sec7}.

\section{Theoretical predictions versus observations}
\label{sec2}

Standard CV population models predict that $\sim90-99$ per cent of the systems should be below the orbital period gap 
and that $\sim40-70$ per cent should have brown dwarf donors (known as ``period bouncers'') 
with a strong accumulation of systems around the orbital period minimum \citep[e.g.][]{kolb93-1,howelletal01-1,goliasch+nelson15-1}.
This is a natural consequence of the abrupt decrease in the rate of mass transfer when the secondary star becomes fully convective 
and magnetic braking stops (or at least becomes inefficient) at the upper edge of the gap, predicted by the disrupted magnetic braking model.
Mass transfer rates below the gap are $1-2$ orders of magnitude lower than above, and decrease progressively as the donor loses mass.
This implies that the systems must spend most of their lives below the gap and especially around the minimum orbital period,
where mass transfer proceeds at a very low rate.

Observationally, the number of CVs above and below the gap was roughly the same for decades \citep{ritter+kolb03-1}, 
and the predicted accumulation of systems towards the shortest periods, as well as the high fraction of period bouncers, were
also not detected \citep[e.g.][]{patterson98-1,knigge06-1}.
This has been attributed to observational biases, because systems below the gap would be much fainter, especially at the shorter 
orbital periods. Indeed, this bias was partially overcome in the last decade when a deep survey of CVs from the Sloan Digital 
Sky Survey (SDSS) spectroscopic database \citep{szkodyetal02-2,szkodyetal03-2,szkodyetal04-1,szkodyetal06-1,szkodyetal07-2,szkodyetal09-1,szkodyetal11-1}
allowed the identification of faint CVs previously undetected.
This survey revealed a much larger fraction of CVs with short periods compared with any previous CV sample and an accumulation of 
systems close to the shortest periods \citep{gaensickeetal09-2}.
In addition, \citet{mcallisteretal19-1} recently estimated donor star masses for a sample of 225 CVs showing superhump phenomena \citep{papaloizou+pringle79-1}
and found that 30 per cent are likely to be period bouncers with brown dwarf donors. However, \citet{palaetal19-1} found a fraction of only
5 per cent of period bouncers in a volume limited sample of CVs.
Although the fraction of observed systems below the gap did not yet reached the numbers predicted by models, at least a larger 
fraction of systems is now being detected below than above the gap, and it seems that the previously missing spike 
at the orbital period minimum is finally being detected. 

The location of the orbital period minimum has been another discrepancy between models and observations. 
Theoretical simulations where gravitational radiation is assumed to be the only angular momentum
loss mechanism for fully convective donor stars predict a minimum orbital period of $\sim 65-70$\,min 
\citep{kolb93-1,kolb+baraffe99-1,howelletal01-1,goliasch+nelson15-1,kalomenietal16-1}, while the observed value is $\sim 76-82$\,min 
\citep{knigge06-1,kniggeetal11-1,mcallisteretal19-1}.
However, the simulated location of the orbital period minimum could be shifted towards larger periods if an additional source of 
angular momentum loss, apart from gravitational radiation, is present below the gap \citep{kolb+baraffe99-1,kniggeetal11-1}.
Observationally, the measured radii of donors in CVs \citep{patterson98-1,kniggeetal11-1} as well as recent measurements of accurate 
effective temperature of WDs in CVs \citep{palaetal17-1} seem to support the idea that angular momentum loss below the gap is stronger 
than that caused by gravitational radiation alone. 
Recently, \citet{mcallisteretal19-1} found that the observed donor properties of CVs are consistent with the standard disrupted magnetic 
braking model for systems immediately below the period gap, but an additional source of angular momentum loss is needed at shorter orbital periods.
However, the origin of this extra angular momentum loss is still not clear, as we will discuss in more detail in the Section\,\ref{sec6}.

Finally, the predicted space density of CVs derived from simulations that assume the standard CV evolutionary model 
\citep[e.g.][]{ritter+burkert86-1,dekool92-1,kolb93-1} exceeds those derived from observations by $1-2$ orders of magnitude 
\citep[e.g.][]{patterson98-1,schreiber+gaensicke03-1,pretorius+knigge12-1,brittetal15-1,palaetal19-1}.

\section{The WD mass problem}
\label{sec3}

While the recent observational advances that allow the identification of faint CVs helped to solve, or at least improve, 
the main disagreements between models and observations of CVs that existed for decades, they also revealed a new problem 
which probably became the most serious one during the last decade.

The progenitors of CVs, i.e. close but detached white dwarf plus main sequence (WD + MS) binaries, are assumed to descend from relatively wide MS + MS binaries,
in which the more massive star becomes a giant and fills its Roche-lobe starting to transfer mass to its MS companion. 
Mass transfer under this conditions is dynamically unstable and the transferred mass forms a common envelope that 
surrounds both the core of the primary (the future WD) and the secondary, whose separation is dramatically reduced in 
a spiraling-in process due to drag forces within the envelope. Angular momentum and orbital energy are transferred from the 
binary orbit to the envelope which is finally expelled leaving behind a close WD + MS binary \citep{paczynski76-1}. 
The reduction of the orbital separation, and therefore the probability of the system to avoid a merge during this stage, 
depends on the efficiency at which the energy extracted from the orbit is used to eject the envelope, which is 
known as the common-envelope efficiency \citep[e.g.,][]{zorotovicetal10-1}. 

As a consequence of this evolution, the average WD mass in CVs and in their detached WD + MS progenitors should be smaller than the mean WD mass of 
single stars (i.e. $\sim0.6\Msun$), because the presence of the low-mass companion helps to eject the envelope of the giant primary
prematurely, terminating the mass growth of its core. There must also be a considerable fraction of systems with low-mass 
helium-core WDs ($\Mwd\,\lappr\,0.5\Msun$) descending from systems in which the primary filled its Roche lobe on the first giant branch,
which are not expected to exist among single WDs descending from single stars, because their evolutionary timescale exceeds the age 
of the Galaxy.
Indeed, early binary population models of CVs and their progenitors predict a mean WD mass of $\sim0.5\,\Msun$, well below the mean WD 
mass of single stars, and a large fraction of systems with helium-core WDs that can make up to $\sim50\%$ 
depending on assumptions like the initial-mass-ratio distributions and the common-envelope efficiency
\citep{dekool92-1,dekool+ritter93-1,kolb93-1,kingetal94-1,politano96-1,howelletal01-1}. 

Observations of the direct progenitors of CVs, i.e. detached post-common-envelope WD + MS binaries, are consistent with these
predictions. The average WD mass is similar but smaller than in single WDs
and approximately one third of the systems have a low-mass WD primary \citep[e.g.][]{zorotovicetal11-1,zorotovicetal11-2},
which is consistent with simulations that assume a small common-envelope efficiency \citep[e.g.,][]{zorotovicetal14-1}. 

The picture is completely different for CVs. Since the first measurements, the observed WD masses have been significantly higher than predicted, 
with most of them in the range of $0.8-1.2\,\Msun$ \citep[e.g.,][]{warner73-1, warner76-1, ritter76-1, robinson76-1}, which has been interpreted as a 
selection effect. Basically, old CV samples are dominated by bright systems, discovered because of the accretion generated 
luminosity. CVs with larger WD masses release more energy per accreted unit mass and have more extended accretion disks around the WD, being on average 
significantly brighter and easier to be discovered. Therefore, magnitude-limited CV samples are biased against low-mass WDs, but the 
effect should nearly disappear for fainter CVs \citep{ritter+burkert86-1}. 
However, as we already mentioned, a deep survey with faint CVs from the SDSS is now available \citep[][]{gaensickeetal09-2} and the observed WD masses 
are still too high \citep[e.g.,][]{littlefairetal08-1,savouryetal11-1,zorotovicetal11-1}. 
The old explanation of an observational bias is no longer valid for these systems, because the optical emission in faint CVs is dominated by the
luminosity of the WD instead of the accretion disk. The probability of identifying CVs within SDSS has been calculated in \citet{zorotovicetal11-1} for systems 
dominated by the luminosity of the WD, taking into account the WD effective temperature, WD radius, absolute i-band magnitude, minimum and maximum 
distance at which SDSS will have obtained follow-up spectroscopy and effective survey volume. 
The conclusion is that SDSS CVs should be biased against systems with high-mass WDs, mainly due to the WD mass-radius relation, and that low-mass
helium-core WDs, if they exist, should be more easily detected. Combining the systems identified within SDSS with previously known CVs, 
the observed average mass of the WD is $0.82\Msun$ for the whole sample and $0.83\Msun$ for a sub-sample of CVs with the more confident WD mass 
estimations (mostly eclipsing systems). This value is consistent with the average WD mass derived in the most recent surveys \citep{mcallisteretal19-1,palaetal19-1}.
CVs with confirmed helium-core WDs are still absent, even in surveys that favor their 
detection. These results challenged the standard theory for CV evolution, which had been used for decades for their simulation. 
The observed disagreements could not be due to our poor understanding of the common envelope phase, because the average WD mass and the fraction of
helium-core WDs in detached post-common-envelope binaries are consistent with the theoretical predictions. 
So there must be something that we were not understanding in the evolution of CVs, which should happen after the common-envelope phase, 
causing systems with low-mass WDs to disappear from CV samples, either because the WD mass grows or because  
the systems merge. Any of the above meant that the standard evolutionary model of CVs had missing ingredients
\citep[see][for a detailed discussion of the WD mass problem]{zorotovicetal11-1}.

Mass growth of the WD during the CV phase or in a preceding short phase of thermal time-scale mass transfer for many CVs was investigated in 
detail in \citet{wijnenetal15-1}, concluding that both options can be ruled out as the main cause for the WD mass problem \citep[see also][]{liu+li16-1}. 
By plotting the predicted CV population in a mass ratio ($q=\Msec/\Mwd$) versus secondary mass ($\Msec$) diagram it became evident
that mass growth was not the solution \citep{schreiberetal16-1}. Models predicted too many CVs with low-mass WDs and very low-mass secondary 
stars. Even if one assumes that ALL the mass of the donor star is used to increase the mass of the WD, the observed average mass would not be 
reached and a significant fraction of helium-core WDs would remain.

\section{Consequential angular momentum loss}
\label{sec4}

A critical parameter for population models of CVs is the angular momentum loss rate, which affects the limit in mass-ratio for stability 
against dynamical time-scale mass transfer.
The stability limit is derived by equating the adiabatic mass-radius exponent ($\zeta_{\mathrm{ad}}$) with the mass-radius exponent 
of the donors Roche-lobe ($\zeta_{\mathrm{RL}}$):
\begin{equation}
\zeta_{\mathrm{ad}}=\left(\frac{dln\Rsec}{dln\Msec}\right)_{ad}=\left(\frac{dln\Rsec}{dln\Msec}\right)=\zeta_{\mathrm{RL}}. 
\end{equation}
The latter depends strongly on angular momentum loss. 

Fully-conservative models assume conservation of both angular momentum and the total mass of the binary. 
In this case, only ``systemic'' angular momentum loss through magnetic braking and gravitational radiation is considered, 
which is independent of mass transfer. The mass-radius exponent of the donors Roche-lobe in this model is:
\begin{equation}
\zeta_{\mathrm{RL}} = \frac{2}{3}\frac{ln(1+q^{1/3})-\frac{1}{2}\frac{q^{1/3}}{(1+q^{1/3})}}{0.6q^{2/3}+ln(1+q^{1/3})}(1+q)+2(q-1),
\end{equation} 
where $q=\Msec/\Mwd$ is the mass ratio of the system. 

However, CVs experienced nova eruptions in which they lose the mass that has been previously accreted and 
which is often ignored but must be present in the equation. Mass loss should cause an extra source of angular momentum loss, which is 
known as ``consequential angular momentum loss'' (CAML) because it is a direct consequence of mass transfer \citep[e.g.,][]{king+kolb95-1}.
According to \citet{schenkeretal98-1}, CAML can be incorporated as a continuous effect in the angular momentum loss rate:
\begin{equation}
\Jdot = \Jsysdot + \Jcamldot,
\end{equation}
where $\Jsysdot$ is the systemic angular momentum loss (through magnetic braking and gravitational radiation), present even in 
the absence of mass transfer, and $\Jcamldot$ is the angular momentum loss caused by nova eruptions, which can be derived from: 
\begin{equation}
\frac{\Jcamldot}{J} = \nu \frac{\Msecdot}{\Msec},
\end{equation}
where $\nu$ might depend on the masses of the two stars. This translates into a mass-radius exponent for the donors Roche-lobe of:
\begin{equation}
\zeta_{\mathrm{RL}} = \frac{2}{3}\left(\frac{ln(1+q^{1/3})-\frac{1}{2}\frac{q^{1/3}}{(1+q^{1/3})}}{0.6q^{2/3}+ln(1+q^{1/3})}\right)+2\nu+\frac{\Msec}{(\Mwd+\Msec)}-2.
\end{equation}

The classical non-conservative model for CAML from \citet{king+kolb95-1}
assumes that the ejected mass takes away the specific angular momentum of the WD, which implies:
\begin{equation}
\nu = \frac{\Msec^2}{\Mwd(\Mwd+\Msec)}.
\end{equation}

Recently, \citet{schreiberetal16-1} simulated the predicted population of CVs for both, the fully-conservative model and the model that includes
the classical non-conservative CAML from \citet{king+kolb95-1}. The disagreement between predictions and observations 
was even worse for models that include the classical form of CAML, which predicts more systems below the gap (with $\Msec\,\lappr\,0.35\Msun$)
and a larger fraction of CVs with helium-core WDs. This is because the classical form of CAML brings the system closer, but the 
effect is compensated by the mass loss of the WD, which significantly reduces the decrease of Roche lobe of the secondary star
increasing the limits in the mass-ratio for stable mass transfer. 
It became evident from this simulations that the WD mass problem could be solve if there is a source of angular momentum loss
that drives especially CVs with low-mass WDs into dynamically unstable mass transfer.

The stability limits for mass transfer in CVs where therefore reviewed in \citet{schreiberetal16-1}, by incorporating an equation for 
CAML that is inversely proportional to the WD mass and independent of the mass of the secondary star, i.e. where the specific angular
momentum of the lost matter increases with decreasing WD mass. This translates into:
\begin{equation}
\nu = \frac{C}{\Mwd},
\end{equation}
where C is a constant that we derived to be $C\,\sim\,0.37$ to better fit the observed WD mass distribution.
This is know as the ``empirical CAML model'', because it was empirically derived from the observations. 

\begin{figure*}
\includegraphics[angle=270,width=\textwidth]{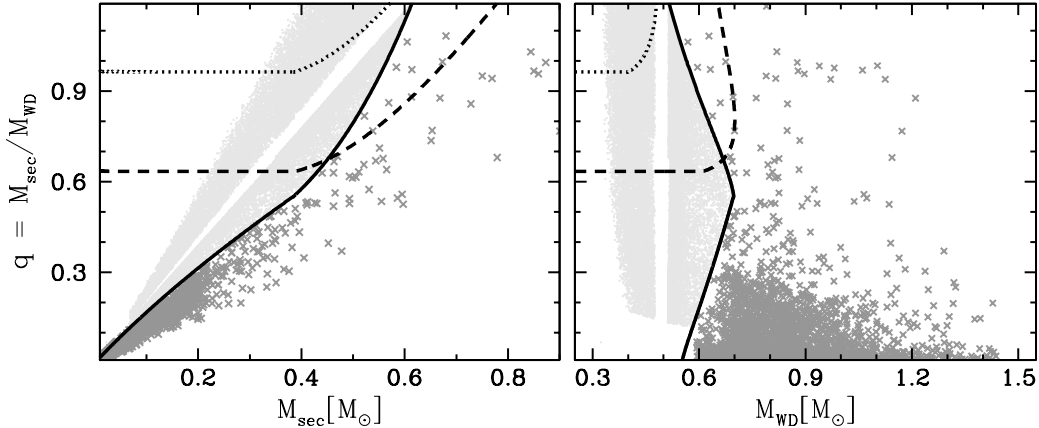}
\caption{Predicted CV population (gray crosses) in the q versus $\Msec$ diagram (left) and q versus $\Mwd$ diagram (right).
The lines represent the critical mass-ratio for the fully-conservative case (dashed line), the classical non-conservative model for CAML
(dotted line), and the empirical CAML model (solid line). The lighter gray crosses correspond to systems that merge according to the 
empirical model from \citet{schreiberetal16-1}.
Figure adapted from \citet{zorotovic+schreiber17-1} for this review.}
\label{figq}
\end{figure*}

Figure\,\ref{figq} shows the mass ratio versus secondary star mass (left) and WD mass (right) for the predicted population of CVs 
from \citet{zorotovic+schreiber17-1}. The lines correspond to the limits in mass ratio above which mass transfer
is dynamically unstable according to the three models explained here: the fully-conservative model (dashed line), 
the classical non-conservative model for CAML from \citet[][dotted line]{king+kolb95-1}, and the empirical CAML model
from \citet[][solid line]{schreiberetal16-1}. Systems that should merge according to the empirical model 
are shown in light gray. 

\section{Testing the new empirical model}
\label{sec5}

The WD mass distribution of CVs can be successfully reproduced by the model developed by \citet{schreiberetal16-1}. 
This is not a surprise, since the model was empirically adjusted for this purpose. 
However, in this section we will show how the predictions of this model fit also better with other observational
properties of CVs and even for detached post-common-envelope WD + MS binaries and single WDs.

\subsection{Orbital period distribution}

As shown in \citet{schreiberetal16-1}, simulations with the empirical model for CAML predict a larger fraction of 
systems above the orbital period gap compared to the predictions of previous models. Although this new model still 
predicts that most systems must be below the gap, which deeper surveys of CVs revealed to be true,
the simulated distribution of orbital periods produced by this model is in better agreement with the observations.

\subsection{Location of the minimum period}

As we mentioned in Section\,\ref{sec2}, the location of the orbital period minimum is observed to be at a larger period than
predicted by simulations if gravitational radiation is assumed to be the only source of angular momentum loss for 
systems with low-mass donors. This suggests that there must be an extra source of angular momentum loss below the gap
\citep{patterson98-1,kniggeetal11-1} and CAML seems to be a good candidate for solving the problem.

\subsection{Space density of CVs}

The new stability limit derived from the empirical model could also solve the space density problem. In \citet{bellonietal18-1} we 
performed binary population synthesis of CVs and concluded that the CV space density predicted
by the empirical model for CAML \citep{schreiberetal16-1} is considerably smaller than the value 
predicted by the classical non-conservative model \citep{king+kolb95-1}, because of the larger fraction of
systems that merge instead of becoming CV. 

\subsection{Detached CVs in the period gap}
A direct test for the disrupted magnetic braking model that allows to explain the orbital period gap in CVs
can be obtained from the observation of detached WD + MS binaries, as proposed by \citet{davisetal08-1}. 
Basically, if CVs stop mass transfer at the upper edge of the gap and become detached, the deficit of CVs 
observed in the period range of $\simeq\,2-3$\,h must be translated into an increase of detached systems 
in the same period range. 

According to \citet{zorotovicetal16-1}, CVs that became detached systems and are crossing the period gap 
without mass transfer should have secondary stars in the spectral-type range of M4 to M6. 
Identifying a peak in the population of observed detached systems with secondaries in this spectral-type range 
at the location of the period gap would not only prove that CVs are crossing the gap without mass transfer, 
but would also allow to test the empirical CAML model from \citet{schreiberetal16-1}. The CVs predicted by 
this model are on average more massive than older predictions, and therefore the detached systems in 
this period range should also have a larger fraction of massive WDs, if a substantial fraction of 
them correspond to previous CVs that became detached. 

Indeed, the expected peak at $\Porb\,\simeq\,2-3$\,h was observed in the period distribution of detached WD + MS systems 
with secondaries in the spectral-type range of M4 to M6. 
This peak could not be reproduced without the inclusion of detached CVs in the simulations, which was interpreted by 
\citet{zorotovicetal16-1} as direct evidence for the disrupted magnetic braking theory. 

Based on the distribution of periods only, it was not possible to statistically favor either of the two models for CAML (classical or empirical),
since both were consistent with the observations given the low number of observed systems. 
However, a large fraction of systems with massive WDs was observed in the location of the period gap, which is better reproduced by the 
empirical model for CAML proposed by \citet{schreiberetal16-1}.

\subsection{Single low-mass WDs}

Finally, the empirical model recently derived for the evolution of CVs has a direct impact on the simulated population of single WDs. 
This model predicts a much larger fraction of CV mergers, especially in systems with low-mass WDs, compared to previous models. 
This translates into a larger number of single WDs as a result of the merger process and it is therefore a potential explanation 
for the existence of single low-mass (helium-core) WDs. 

Low-mass WDs, with $\Mwd\,\lappr\,0.5\Msun$, cannot descend from single stars, since the Universe is not old enough for their progenitors to evolve.
Indeed, there is an estimated fraction of observed low-mass WDs of $\sim\,10$ per cent in the solar neighborhood \citep{kepleretal07-1} and 
the vast majority of them belong to a close binary system \citep{marshetal95-1,brownetal10-1,rebassa-mansergasetal11-1}.
However, $\lappr\,20-30$ per cent of the low-mass WDs seem to be single \citep{brownetal11-1}, which translates into $\lappr\,2-3$ per cent of
single low-mass WDs among the whole population of observed WDs.

Several explanations have been proposed to explain the existence of single low-mass WDs: merger events either during common-envelope evolution 
or of two very low-mass WDs \citep{nelemans10-1,brownetal11-1}; sub-stellar companions that 
help to eject the envelope and get evaporated or suffer a merge \citep{nelemans+tauris98-1}; strong mass loss in metal-rich stars close to the tip 
of the first giant branch \citep{kilicetal07-1,brownetal11-1,hanetal94-1,mengetal08-1}; remnants of the companion stars in type Ia Supernovae 
\citep{justhametal09-1,wang+han09-1}. 
However, none of this channels has been able to convincingly reproduce the observed fraction so far \citep[see][for a review]{zorotovic+schreiber17-1}.

In \citet{zorotovic+schreiber17-1}, we simulated the population of single WDs, taking into account systems that evolve from single stars 
and from different merger channels in binary stars, including (for the first time) CV mergers predicted by the empirical model for CAML. 
The predicted relative number of low-mass WDs among single WDs was consistent with observations, with merging CVs being the dominant channel leading to 
their formation. 

\section{Origin of extra angular momentum loss}
\label{sec6}

As we have reviewed in Sections\,\ref{sec4} and \ref{sec5}, the empirical model for CAML derived by \citet{schreiberetal16-1} to solve 
the WD mass problem in CVs can solve at the same time several disagreements between observations and simulations of CVs that have existed
for decades, and it has also been successfully tested with observations of detached WD + MS systems and even single WDs.

However, the model is still purely empirical, and the physical origin of the dependence of CAML on the WD mass
must be understood in order to progress in our understanding of CV evolution.

Different mechanisms for CAML proposed in the literature were recently reviewed by \citet{shao+li12-1}, including
the formation of a circumbinary disk that can drain orbital angular momentum from the system \citep{vandenheuvel94-1,taam+spruit01-1},
an isotropic wind from the WD surface \citep{king+kolb95-1}, and outflows of mass through the inner or outer Lagrangian points \citep{vanbeverenetal98-1}.
They concluded that a wind from the WD surface or the loss of mass through the inner Lagrangian point cannot account for the extra angular momentum loss derived 
from observations of CVs below the period gap \citep{kniggeetal11-1}. A circumbinary disk or outflows from an outer Lagrangian point are able to account for it 
if a considerable fraction of the mass transferred from the donor to the WD leaves the system. 

A promising candidate for CAML that increases with decreasing WD mass is angular momentum loss due to friction 
between the secondary star and the expanding nova shell in a nova eruption \citep{schenkeretal98-1,schreiberetal16-1,nelemansetal16-1}. 
The specific angular momentum that can be extracted by friction depends strongly, and inversely, on the expansion velocity of the ejecta
at the location of the donor, and it can be very high for slow ejections \citep{schenkeretal98-1}. On the other hand, the maximum velocity of the ejecta is much larger
for massive WDs \citep{yaronetal05-1}. This would imply that, during a nova eruption, the systems with less massive WDs eject the material 
at smaller velocities, experiencing more angular momentum loss by friction than systems with massive WDs, which makes it harder for them 
to remain within the regime of stable mass transfer. In other words, CVs with low-mass WDs could form, but they would quickly 
become unstable, probably even after the first nova eruption. 
According to recent simulations from \citet{liu+li19-1}, in order to account for the extra angular momentum loss (apart from gravitational radiation) 
for systems below the orbital period gap, the velocities of expansion of the material ejected in a nova outburst should be extremely 
low, resembling a common-envelope phase. Indeed, \citet{nelemansetal16-1}
found that an ejection similar to a common envelope might significantly affect the stability of mass transfer, 
particularly for systems with low-mass WDs. 
In another approach, \citet{liu+li16-1} calculated that if $\sim20-30$ per cent of the matter ejected during a nova eruption
remains in the system forming a circumbinary disk, CVs with low-mass WDs would be more likely to become dynamically unstable.

As can be seen, several studies have tried to understand the possible forms of CAML that affect CV evolution in recent years and, 
although progress has been made, it remains an open field for research.  

\section{Summary}
\label{sec7}

We have summarized the main disagreements between the theory and the observations of CVs that have existed for decades, 
and their possible solutions. In addition, we have shown how a deeper survey of CVs has revealed a new problem during the 
last decade, related to the mass of the WD and the absence of low-mass helium-core WDs. 

We explained the basics of a new model for the evolution of CVs proposed by 
\citet{schreiberetal16-1} which assumes enhanced consequential angular momentum loss for systems with low-mass WDs.
This model does not only solve the WD mass problem, for which it was empirically adjusted, 
but also fits better with the observations of the orbital period distribution, the location of the minimum period, and the space 
density of CVs. In addition, this model is in better agreement with observation of detached systems in the period range of $\simeq\,2-3$\,h, 
where detached CVs should be found according to the disrupted magnetic braking model,
and provides a new channel for the formation of single low-mass WDs, which might be the dominant channel for their existence,
bringing into agreement the simulated and observed fractions of low-mass WDs among single WDs. 

The new empirical model for CVs evolution still lacks a physical formulation for the 
origin of the dependence of angular momentum loss on the WD mass that is needed to reproduce the observations. 
However, progress has been made in this area in recent years, and 
promising candidates for angular momentum loss that increases with decreasing WD mass are friction during nova eruptions, possibly resembling 
a common envelope phase, or the formation of a circumbinary disk from material that does not leave the system after the eruption.

\section*{Acknowledgments}
MZ acknowledges support from CONICYT PAI (Concurso Nacional de Inserci\'on en la Academia 2017, Folio 79170121) and CONICYT/FONDECYT (Programa de Iniciaci\'on, Folio 11170559).
MRS acknowledges financial support from FONDECYT Grant No. 1181404. 


\end{document}